# Fairness across Network Positions in Cyberbullying Detection Algorithms

Vivek Singh and Connor Hofenbitzer,
Rutgers University
{vs451, cfh41}@scarletmail.rutgers.edu

*Abstract*— **Cyberbullying, which often has a deeply negative impact on the victim, has grown as a serious issue in Online Social Networks. Recently, researchers have created automated machine learning algorithms to detect Cyberbullying using social and textual features. However, the very algorithms that are intended to fight off one threat (*cyberbullying*) may inadvertently be falling prey to another important threat (*bias* of the automatic detection algorithms). This is exacerbated by the fact that while the current literature on algorithmic fairness has multiple empirical results, metrics, and algorithms for countering bias across immediately observable demographic characteristics (e.g. age, race, gender), there have been no efforts at empirically quantifying the variation in algorithmic performance based on the network role or position of individuals. We audit an existing cyberbullying algorithm using Twitter data for disparity in detection performance based on the network centrality of the potential victim and then demonstrate how this disparity can be countered using an Equalized Odds post-processing technique. The results pave way for more accurate and fair cyberbullying detection algorithms.**

## I. INTRODUCTION

In multiple domains, ranging from automatic face detection to automated decisions on parole, machine learning algorithms have been found to be systematically biased and favoring one demographic group over another [1,2,3]. This is problematic as these algorithms are reinforcing and amplifying existing disparities across different groups of individuals. As a result, certain groups of people may systematically get lesser access to loans, college admissions, parole opportunities, and so on.

At the same time, the discussions around fairness (like in the scenarios above) typically rest on the notion of individual. However, much of the data being produced and the decisions being made today occur in a networked setting. Yet, our social and judicial models are largely centered around the individual. As argued by Boyd et al. [4], we must rethink our models of discrimination and our mechanisms of accountability. We need to look beyond immutable characteristics of individuals and also attend to the positions of individuals in networks.

Hence, understanding the role played by one's position in a network in the quality of decisions one gets from computational algorithms is urgent and important. This work focuses on the fairness of cyberbullying detection algorithms across recipients with different network characteristics or positions. *If the algorithms work quite accurately when an individual with high network centrality is the potential victim and poorly when an individual with low network centrality is the potential victim, then that would be unfair and would reify existing disparities in networks.* In particular, the individuals with lower network centrality will suffer from a "double whammy" because: (1) historical research has shown that individuals on the peripheries of the network tend to be bullied more often those in the center [5]; (2) those in the center of the network tend to have more data available for learning opportunities for the various machine learning algorithms. Hence, algorithms are more likely to work better for those cases where the potential victims are in the center of the network rather than those on the peripheries.

The main contributions of this work are:

*(1) To motivate and ground the use of an individual's network centrality as a sensitive attribute for discrimination analysis. (2) To audit an existing social network features based cyberbullying detection algorithm for bias based on recipient's network position and demonstrate a way to counter it.*

## II. RELATED WORK

There have been multiple recent efforts aimed at increasing the fairness of machine learning algorithms. These approaches can broadly be classified into those that involve *pre-processing* the data going into the algorithms, those that modify the *processing* i.e. prediction algorithm itself, and those that *post-process* the results of an existing algorithm to allow for fairer decisions [1,2,3]. For instance, Calmon et al., propose a pre-processing approach which changes the data going into the algorithms in such a way that tries to maintain the utility at prediction while reducing the dependence of the features on the sensitive attribute (e.g. gender) [1]. Kamishima et al., on the other hand propose adapting the classification algorithms by adding a "regularizer" that penalizes the algorithms for disparate results across considered groups [2]. Hardt et al. proposed a post-processing framework to remove discrimination against a sensitive attribute to predict the target based on available features [3].

None of these fairness-based efforts have focused on network position of a person to identify the favored and disfavored groups. Per our knowledge, ours is the first systematic effort that tackles the issue of fairness based on network position of an affected individual. The two closest related lines of works are [4] and [6]. Boyd et al., [4] argue conceptually about the roles of networks in creating biases but do not deal with any empirical data. Fish et al., [6] on the other hand study the problem of equal access to information as it spreads in a network but studies the problem of "social welfare function", which is very different from the idea of fairness for individuals or groups when considering their specific characteristics. Per our knowledge, ours is the first work that considers network characteristics (e.g. centrality) as a sensitive attribute based on which different emergent groups need to be compared and the disparities countered.

The problem of cyberbullying detection has been studied in multiple domains. Dinakar et al., [7] describe cyberbullying as "when the Internet, cell phones or other devices are used to send or post text or images intended to hurt or embarrass another person." Clearly, cyberbullying involves a content (text, image) component and a social component. However, most of the work on cyberbullying detection focuses on (sophisticated) textual analysis. Work by Huang et al. [8] was the first effort to identify the use of social features in cyberbullying detection. Since then, multiple other efforts (e.g., [9,10]) have also used social features for cyberbullying detection. Given the importance of social aspects (e.g. the network connecting the sender i.e. potential bully and the recipient i.e. potential victim) in cyberbullying, it is important to understand the question of fairness in terms of the network position of the recipient i.e. the potential victim.

To quantify the "fairness" of algorithms, we survey the recent literature on fairness in machine learning (e.g., [1,2,3]) and focus on the comparisons based on three different metrics: difference in accuracy (or AUCROC), equal odds, and equality of opportunity. Equality of opportunity (EoO) metric mandates an equal true positive rate (TPR) for the groups considered (e.g. male and female; or Low network centrality and High network centrality). Almost all practical algorithms have TPR below 100%. In such cases, EoO principle mandates the difference of TPR for the considered groups should be as low as possible. In other words, a ground truth based "true" cyberbullying post should have equal odds of being labeled as "true" for cyberbullying by the detection algorithm irrespective of the network centrality of the recipient (potential victim). Equal odds metric is an extension of the above idea to include both the true positive rate and the false positive rate [3]. Hence, the difference in the false positive rates for different considered groups (e.g. low/high centrality) is also considered in this work. The overall goal of this work is to minimize the discrepancy in the accuracy, TPR, and FPR based on the centrality of the message recipient.

### III. DATASET AND APPROACH

Like our previous work [8], the social and textual features have been defined in this work as follows. The social features were derived using the 1.5 ego-networks, where 'ego' refers to an individual focal node. Let the 'global social network' be represented as a graph $G = <V; E>$ where $V$ is the set of all the nodes and $E$ is the set of directed edges over those nodes. The 1-ego network of a node $v$ is defined as the graph $G_1(V_1; E_1)$ such that it includes all the neighbors of $v$. The 1.5-ego-network is defined as the graph $G_{1.5}(V_{1.5}; E_{1.5})$ such that it includes the all the interconnections (edges) between the nodes present in the 1-ego-network defined above.

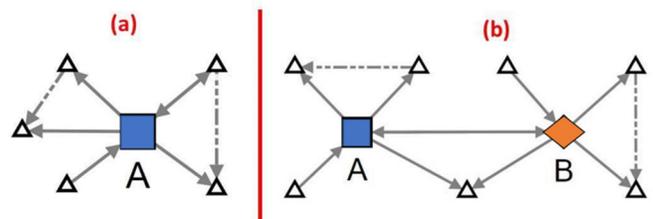

Figure 1: Schematic representation of 1.5 ego-network relationship graphs used to derive network features.

In Fig. 1(a), the ego-node $A$ has been shown as a solid blue square and the neighbors of $A$ are marked as triangles. The solid lines represent the edges of the 1-ego-network i.e. $E_1$ while the dashed lines represent the additional edges in $E_{1.5}$. In this work, we focus on 1.5-ego-networks to describe an individual's social network as they capture social context at a reasonable level (the focused node, their friends, and the relationships between those friends) while keeping the data requirements and computational complexity low [8].

As shown in Fig. 1(b), we define the relationship graph of two users, with a sender, defined '$A$' and a receiver defined '$B$' (shown as an orange trapezoid) by combining the 1.5-ego-networks of the two users. These relationship graphs allow one to describe the sender and receiver nodes in terms of social activity. For example, the relationship graph can be used to describe which users are central to a network and characterize which friends the two parties share.

Specifically, similar to [8], the following social network features are defined for the relationship graph: (1) number of nodes, (2) number of edges, (3) degree centrality- with variants for in-degree, out-degree, sender and receiver resulting in four different features, (4) edge betweenness centrality of the edge between sender and receiver, (5) tie strength between a sender and a receiver, and (6) community embeddedness measured as a k-core score for the sender and receiver (two features) [8], resulting in a total of ten social features to describe a user's social interactions.

Similarly, based on [8] the following textual features were included in the modeling: (1) density of bad words, (2) density of uppercase letters, (3) number of exclamation points and question marks, (4) number of smileys, and (5) part-of-speech-tags, these were chosen based on their correlation to the predictors output.

We use a labeled cyberbullying dataset as utilized in [8]. This dataset is a subset of the Twitter corpus from the CAW 2.0 data set, which has been annotated by three labelers for the magnitude of cyberbullying. This data set contains 4,865 messages with 93 (roughly 2%) of them labeled as bullying messages. One of the largest problems with cyberbullying datasets is the data imbalance. To mitigate the effects of imbalance, we applied the 'SMOTE' method [11]. SMOTE (Synthetic Minority Oversampling TEchnique) this works by under sampling the majority class and over sampling the minority class. However, it mitigates the problem of overfitting caused by simple replication of data points by generating newer (synthetic) examples by operating in 'feature space' rather than 'data space' [11].

To train and validate the predictions we conducted a 70%-30% split after shuffling the dataset to allow for instances of cyberbullying to be in both the training and testing set. We then applied SMOTE preprocessing on the training set. This resulted in an equal number of bullying and non-bullying instances and increased the number of instances in the training set from 3,420 to around 6,750. This allowed for more instances of the minority class to be used in training, potentially increasing the accuracy of predictions. The test set remains imbalanced in the ratio of 98:02 as indicated above to mimic the real-world scenario.

After SMOTE, we applied a dagging (Directed Aggregating) algorithm, which was reported as the best performing algorithm in [8], to create a model for cyberbullying detection. From the dagging predictor we obtained probabilities for the testing set predictions, which allowed us to calculate receiver operating characteristic metrics which were used later for auditing and applying equalized odds post processing. We were able to obtain probability scores by using the notion of soft-voting, which is the average of the models voting rather than a hard cut off for each model.

*A. Auditing algorithm for bias*

To identify a sensitive attribute, we considered features which could be linked to an individual's network position and network activity as these could be unfairly affecting a user's probability of being identified as a target of cyberbullying. Here we use "outdegree centrality for recipient" as a sensitive attribute as the recipient is the person likely to be vulnerable in cyberbullying and the outdegree centrality can give a clue as to how active a person is on the network. This feature also has associations with a person's introversion/extroversion personality trait descriptor, which again goes beyond the traditionally studied focus on immediately discernable characteristics like age, and gender.

As suggested in recent efforts on fair machine learning [1,3], the sensitive attribute was not included in the algorithm's predictions as this could lead to more biases in the predictions. We calculated the median of the sensitive attribute to create two groups– those with "high" network centrality and those with "low" network centrality. Next, we audited the outputs of the algorithm for possible bias. We computed the above-mentioned algorithm's predictions, through which we were able to calculate receiver operating characteristic (AUCROC) scores as well as other performance metrics (TPR, FPR) for the two groups. Note that AUCROC is a more robust metric for measuring the performance of algorithms and is preferred to simple accuracy metric in scenarios involving imbalance across classes [11]. The above process allowed us to determine the difference in accuracy metrics across the two groups.

We ran the auditing algorithm 100 times to allow for more confidence in the average results obtained and to also use the variations in the results to determine statistical significance of the difference. Each test round used a new random seed that was used for the test-train split, meaning that different population samples were being used to train and to test the algorithm's performance in each round. The average results are shown in Table 1.

| Attribute | Baseline | | |
|---|---|---|---|
| | TPR | FPR | ROC AUC |
| "High" network centrality | 0.8102 | 0.3801 | 0.7714 |
| "Low" network centrality | 0.5328 | 0.1398 | 0.7153 |
| Difference | 0.2774 | 0.2403 | 0.0561 |

Table 1: The average TPR, FPR and area under the ROC curve comparison for those with a "low" network centrality and those with a "high" network centrality in the baseline method.

Throughout the analysis, we found that accuracy scores were higher when the recipients of the messages had "high" network centrality than when the recipients had "low" network centrality. For instance, the true positive rate (a very important metric in minority true class scenarios, like cyberbullying) was 0.8102 for the "high" network centrality group and only 0.5328 for the "low" network centrality group. Using the R statistical language, we conducted a *t-test* with alpha=0.05 threshold for TPR, FPR and AUCROC difference between the groups. We found that the difference in prediction accuracy metrics was ***statistically significant*** between the two considered groups.

*B. Debiasing algorithm using equalized odds post-processing*

Equal odds principle requires the TPR and FPR to be equal for both the underprivileged and privileged classes. Using a machine learning algorithm's prediction, accuracy metrics can be calculated at different classification thresholds. Here we adapted the Equalized Odds Post-processing approach as proposed by [3] and as available in the IBM AIF 360 library [12] to compute the receiver operating characteristic for the considered groups. Using the AIF 360 library we implemented the classification metric class to obtain various performance values (AUC ROC, TPR, FPR) for each group *before* and *after* the debiasing process. The library was adapted to include calculations for area under the receiver operating characteristic curve between two groups to better suit this paper as the original library had no notion of AUCROC.

We say that a predictor $\hat{Y}$ has achieved equalized odds [3] with respect to the sensitive/protected attribute A and the outcome Y if the following is satisfied:

$Pr\{\hat{Y}=1 \mid A=0, Y=y\} = Pr\{\hat{Y}=1 \mid A=1, Y=y\}$, $y \in \{0,1\}$

In designing a derived predictor from binary $\tilde{Y}$ and A we can only set four parameters: the conditional probabilities $p_{ya} = Pr\{(\tilde{Y}=1 \mid \hat{Y}=a, A=a\}$. These four parameters, $p = (p_{00}, p_{01}, p_{10}, p_{11})$, together specify the derived predictor $\tilde{Y}_p$. For equal odds, this requires that for the outcome $y$, $\hat{Y}$ has equal positive rates for each group, A = 0, A = 1. Since the expected loss $El(\tilde{Y}_p, Y)$ is also linear in *p*, the optimal derived predictor can be obtained as a solution to the following linear program with four variables and two equality constraints:

$$\min_{p} El(\tilde{Y}_p, Y)$$

$$\text{s.t. } \gamma_0(\widetilde{Y_p}) = \gamma_1(\widetilde{Y_p}) \text{ and}$$
$$\forall_{y,a} \leq p_{ya} \leq 1$$

where the components of $\gamma_a(\widetilde{Y})$ are the false positive rate and the true positive rate within the considered group A = a. The goal of Equalized Odds post-processing is to maximize the accuracy while keeping the difference in metrics between considered groups to a minimum.

## IV. RESULTS

Table 2 shows the results for AUC ROC, TPR (True Positive Rate), and FPR (False Positive Rate) after applying Equalized Odds post processing. The comparison between the two approaches are summarized in Table 3. The results indicate that the proposed approach resulted in lower discrepancy between the two considered groups in terms of TPR and FPR, which works towards the notion of equalized odds. We also notice a decrease in difference of ROC between the two methods. These decreases in differences were validated using one-sided t-tests. The differences in scores for TPR, FPR, and AUC were found to be *statistically significant* at alpha = 0.05 threshold. Please note that this increase in fairness also comes with a slight decrease in overall AUC from 0.7434 to 0.7283, which was found to be *not statistically significant*.

| Attributes | Proposed Method | | |
|---|---|---|---|
| | TPR | FPR | ROC AUC |
| "High" network centrality | 0.7019 | 0.3339 | 0.7112 |
| "Low" network centrality | 0.5379 | 0.1427 | 0.7454 |
| Difference | 0.1641 | 0.1912 | -0.0342 |

Table 2: The average TPR, FPR and area under the ROC curve comparison for those with a "low" network centrality and those with a "high" network centrality after debiasing.

| Attributes | Deltas across high/low centrality groups | | |
|---|---|---|---|
| | TPR | FPR | ROC AUC |
| Baseline |Delta| | 0.2774 | 0.2403 | 0.0561 |
| Proposed |Delta| | 0.1641 | 0.1912 | 0.0342 |
| Change | 0.1133 | 0.0492 | 0.0119 |

Table 3: Comparison of the deltas between the two groups (low-centrality and high-centrality) in the baseline and the proposed approach.

Based on the trends in the considered dataset we find that the proposed approach is useful at reducing the disparity in the performance of cyberbullying detection algorithm across different groups based on the network centrality of the recipients across the metrics of deltas in AUC, TPR and FPR.

The current work also has some limitations. It focuses on a single cyberbullying algorithm applied on a single dataset. The notion of networks considered here focuses on 1.5 ego networks rather than the complete network and a single network feature (outdegree network centrality) has been used an identifier for network position. At the same time, this work marks the first empirical effort at analyzing the difference in performance based on network position of a person – not just in cyberbullying literature but in any application domain. The results obtained here are promising and motivate further work in this direction.

## V. CONCLUSION AND FUTURE WORK

This short paper motivates and grounds the use of network characteristics (e.g. network centrality) as a sensitive attribute to study algorithmic fairness. The audit of a well-cited cyberbullying detection algorithm [8] yielded that the performance of the algorithm varied quite significantly depending on the network centrality of the recipient of the potentially bullying message. This disparity in the performance was found to reduce statistically significantly based on the adoption of the equalized odds post-processing technique. While early, the results significantly move forward the literature on fairness in networked algorithms and specifically cyberbullying detection. Future improvements on this work could consider larger network size, varied representations of network positions, and newer approaches to create fair and accurate network algorithms.